\documentclass[aps,prl,reprint,showpacs,superscriptaddress]{revtex4-1}% Physical Review B
\usepackage{graphicx}% Include figure files
\usepackage{latexsym}
\usepackage[T1]{fontenc}
\usepackage{amsmath}
\usepackage{endnotes}
\input{epsf}

\begin{document}

\preprint{APS/123-QED}

\date{\today}

\title{Lack of coupling between superconductivity and orthorhombic distortion in stoichiometric single-crystalline FeSe}

\author{A. E. Böhmer}
\affiliation{Institut für Festkörperphysik, Karlsruhe Institute for Technology, 76021 Karlsruhe, Germany}
\affiliation{Fakultät für Physik, Karlsruhe Institute for Technology,  76131 Karlsruhe, Germany}

\author{F. Hardy}
\affiliation{Institut für Festkörperphysik, Karlsruhe Institute for Technology, 76021 Karlsruhe, Germany}

\author{F. Eilers}
\affiliation{Institut für Festkörperphysik, Karlsruhe Institute for Technology, 76021 Karlsruhe, Germany}
\affiliation{Fakultät für Physik, Karlsruhe Institute for Technology,  76131 Karlsruhe, Germany}

\author{D. Ernst}
\affiliation{Institut für Festkörperphysik, Karlsruhe Institute for Technology, 76021 Karlsruhe, Germany}

\author{P. Adelmann}
\affiliation{Institut für Festkörperphysik, Karlsruhe Institute for Technology, 76021 Karlsruhe, Germany}

\author{P. Schweiss}
\affiliation{Institut für Festkörperphysik, Karlsruhe Institute for Technology, 76021 Karlsruhe, Germany}

\author{T. Wolf}
\affiliation{Institut für Festkörperphysik, Karlsruhe Institute for Technology, 76021 Karlsruhe, Germany}

\author{C. Meingast}
\affiliation{Institut für Festkörperphysik, Karlsruhe Institute for Technology, 76021 Karlsruhe, Germany}

\begin{abstract}

The coupling between superconductivity and othorhombic distortion is studied in vapor-grown FeSe single crystals using high-resolution thermal-expansion measurements.  In contrast to the Ba122-based (Ba122) superconductors, we find that superconductivity does not reduce the orthorhombicity below $T_c$.  Instead we find that superconductivity couples strongly to the in-plane area, which explains the large hydrostatic pressure effects.  We discuss our results in light of the spin-nematic scenario  and argue that FeSe has many features quite different from the typical Fe-based superconductors. 

%We report on uniaxial pressure effects of vapor-grown $\beta$-FeSe single crystals, studied using high-resolution thermal-expansion measurements. The dramatic increase of the superconducting critical temperature $T_c$ under hydrostatic pressure is demonstrated to arise from a reduction of the in-plane area. In contrast to underdoped 122 pnictides, orthorhombic distortion and superconductivity do not compete in FeSe. Below $\sim 30$ K, non-Fermi-liquid behavior of the thermal expansion is found, which slightly reduces the orthorhombic distortion and presumably arises from low-temperature spin fluctuations. 

%Abstract hat 595 zeichen!!
\end{abstract}

%pnictides and chalcogenides, thermodynamic properties, effects of pressure on the phase diagram

\maketitle

The interplay of structure, magnetism and superconductivity has been a recurrent theme in the study of iron-based superconductors\cite{Stewart2011}. Among these systems, PbO-type $\beta$-FeSe has the simplest crystallographic structure and a rich phase diagram\cite{Stewart2011}. 
It undergoes a structural phase transition, similar to that of many parent compounds of the 1111 and 122 iron-based systems, at $\sim90$ K. In these later materials, structural and magnetic order track each other closely, which has led to the suggestion of a magnetic (spin-nematic) origin of the structural distortion\cite{Fernandes2012}. In FeSe, however, no static magnetism is found at ambient pressure\cite{Bendele2010,McQueen2009} and spin-fluctuations are found to be enhanced only at low temperatures\cite{Imai2009}, which raises the question of the origin of the structural transition.
%At $\sim90$ K, it undergoes a structural phase transition, similar to that of many parent compounds of the 1111 and 122 iron-based systems. In these later materials the structural phase transition is closely tracked by a magnetic transition, and it has been suggested that the structural transition is actually of spin-nematic origin\cite{Fernandes2012}. In FeSe, no static magnetism is found at ambient pressure\cite{Bendele2010,McQueen2009} and spin-fluctuations are found only at low temperatures\cite{Imai2009}, which raises the question of the origin of the structural transition. 
FeSe becomes superconducting at a modest $T_c=8$ K\cite{Hsu2008}, yet the onset of superconductivity rises dramatically to $\sim 37$ K under hydrostatic pressure\cite{Medvedev2009,Margadonna2009}, which also induces static magnetic order\cite{Bendele2012}. Recently, superconductivity even up to over 50 K was demonstrated in strained epitaxial thin films\cite{Wang2012}. 
With its huge sensitivity of $T_c$ to external pressure and the large separation between structurally distorted and magnetically ordered phase, FeSe is an intriguing system to study the phase interplay using pressure as a tuning parameter. Uniaxial pressure effects, as can be studied by thermal expansion, are of special interest, because they are usually very anisotropic in the iron-based systems due to the layered crystal structures\cite{Budko2009,Hardy2009}. 
%However, pressure effects in iron-based systems are usually very anisotropic\cite{..}, due to the layered crystal structures.It is therefore of special interest to also study uniaxial pressure effects in FeSe, as can be done usi high-resolution thermal expansion of single crystals. 
Further, these types of measurements provide a sensitive probe of the coupling between superconducting and orthorhombic order parameters\cite{Meingast2012,Boehmer2012}.
 
Single-crystal growth of $\beta$-FeSe, on the other hand, is complicated by the rich constitutional binary phase diagram of Fe and Se, where the tetragonal, superconducting $\beta$-FeSe phase is located deep below the solidus line\cite{Okamoto1991}. In consequence, crystal growth experiments from the melt or self flux result in hexagonal $\delta$-FeSe, which undergoes a series of structural transformations and decomposition reactions on cooling to room temperature\cite{Okamoto1991}. Only if the initial Se content is low enough, can tetragonal $\beta$-FeSe be obtained, however, only in the form of a platelet of hexagonal morphology which contains both, tetragonal $\beta$-FeSe and magnetic Fe$_7$Se$_8$\cite{Okamoto1991}.

In this Letter, we report on high-resolution thermal-expansion of vapor-grown single-crystalline $\beta$-FeSe. Growth directly in the tetragonal structure results in high quality single-crystals, which are well suited to study the anisotropy of pressure effects; a necessary complement to hydrostatic-pressure studies in any non-cubic material. We show that the dramatic increase of $T_c$ under hydrostatic pressure arises from a reduction of the in-plane area, while $T_c$ is five times less sensitive to the $c$-axis length. Further, we demonstrate non-Fermi-liquid behavior in the low-temperature in-plane thermal-expansion coefficients, which slightly reduces the orthorhombic distortion and presumably arises from low-temperature spin-fluctuations. Surprisingly, orthorhombic distortion and superconductivity do not compete in FeSe, in contrast to underdoped 122 pnictides, which suggests that the structural transition may not be of magnetic origin.  

\begin{figure}[tb]
\includegraphics[width=8cm]{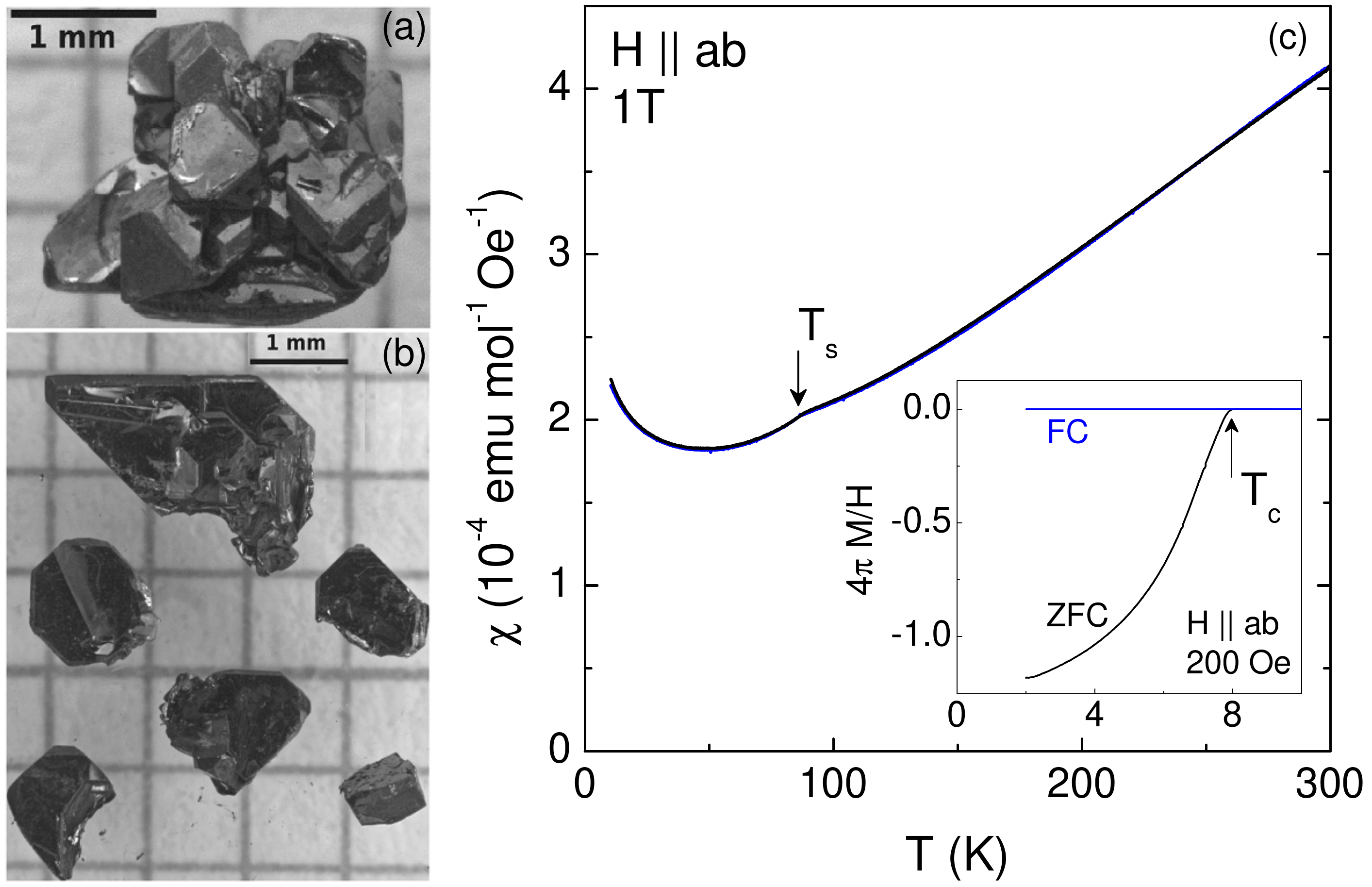}
\caption{(color online) (a,b) Photographs of tetragonal $\beta$-FeSe grown using a low-temperature vapor-transport technique. (c) Temperature dependence of the magnetic susceptibility in a field of 1 T, applied parallel to the $ab$ plane. The inset shows the low-temperature data in a field of 20 mT. The screening is larger than $-1$ because of the demagnetization effect.}
\label{fig:1}
\end{figure}

Fe and Se powders were mixed in an atomic ratio 1.1:1 and sealed in an evacuated SiO$_2$ ampoule together with a eutectic mixture af KCl and AlCl$_3$. The ampoule was  heated to $390^\circ$C on one end while the other end was kept at $240^\circ$C. After 28.5 days isometric FeSe crystals with tetragonal morphology were extracted at the colder end (Fig. \ref{fig:1} (a,b)). At such low temperatures, samples form directly in the tetragonal state and do not undergo structural transformations or decomposition reactions. Wavelength dispersive x-ray spectroscopy reveals an impurity level below 500 ppm, in particular there is no evidence for Cl, Si, K or Al impurities. 

x-ray powder diffraction confirms the tetragonal structure with lattice constants $a=3.7707(12)$ \AA\ and $c=5.521(3)$ \AA. Structural refinement with a 4-circle diffractometer using Mo-radiation yields a composition of Fe:Se=0.995(4):1 (\textit{i. e.} stoichiometric within the error bar) and a structural $z$ parameter of $z=0.26668(9)$. No indications for interstitial atoms were found. Fig. \ref{fig:1} (c) shows the temperature dependence of the magnetic susceptibility with field applied parallel to the $ab$ plane measured in a vibrating sample magnetometer. A small but sharp kink, which we associate with the structural transition, is observed at 87 K. The superconducting transition has a sharp onset at 8 K and is broadened by the relatively high applied field.
High-resolution thermal expansion was measured in a capacitance dilatometer\cite{Meingast1990}, in which the sample is pressed against one plate of a plate-type capacitor with a force of $\sim0.2$ N, directed along the measured sample length. As for underdoped BaFe$_2$As$_2$\cite{Meingast2012,Boehmer2012}, this force can be used to \textit{in-situ} detwin samples of FeSe below their tetragonal-to-orthorhombic phase transition (from space group P4/nmm to Cmma). In particular, if the thermal expansion along $[110]_T$ (tetragonal notation) is measured, samples are detwinned by the applied force and the (shorter) orthorhombic $a$ axis is measured\cite{notenamingab}. Measuring thermal expansion along the tetragonal $[100]_T$ direction yields, ideally, the average of $a$ and $b$ axis. Under this assumption, the thermal expansion of the orthorhombic $b$ axis can be inferred\cite{Boehmer2012}.

\begin{figure}
\includegraphics[width=8cm]{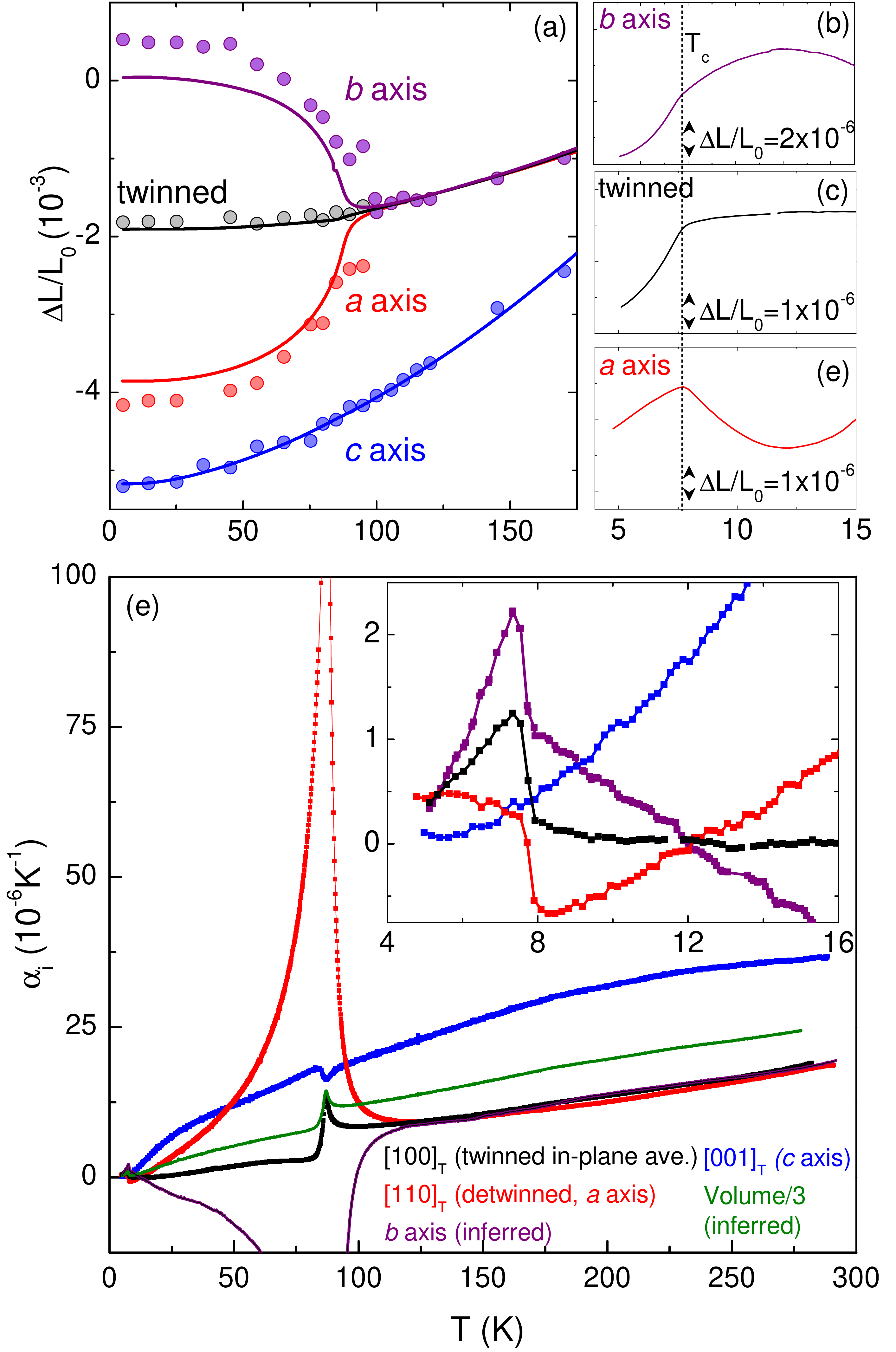}
\caption{(color online) (a) Relative length changes along the three measured directions (orthorhombic $a$ axis, in-plane average and $c$ axis) and inferred $b$-axis length change (continuous lines). For comparison, the corresponding data from x-ray diffraction\cite{Khasanov2010} are given (circles). (b), (c) and (d) show the $b$-axis, in-plane average and $a$-axis length change close to $T_c$, respectively, on a magnified scale. (e) Uniaxial thermal expansion coefficients of FeSe along three measured directions, with the thermal-expansion along the $b$ axis and the volume average inferred from the measurements. The inset shows the data close to $T_c$ on a magnified scale.}
\label{fig:2}
\end{figure}

Fig. \ref{fig:2} shows the thus obtained relative sample-length changes $\frac{\Delta L_i}{L_{i,0}}=\frac{L_i(T)-L_i(300\textnormal{ K})}{L_i(300\textnormal{ K})}$ and uniaxial thermal-expansion coefficients $\alpha_i=\frac{1}{L_i}\frac{dL_i}{dT}$, where the index $i$ stands for the direction. The $\frac{\Delta L_i}{L_{i,0}}$ data are in good agreement with previous x-ray studies\cite{Khasanov2010}  and show clear evidence  for a second-order tetragonal-to-orthorhombic phase transition at $T_s=87$ K. We note that the $c$-axis anomaly at $T_s$ is unusually small when compared to underdoped Ba(Fe,Co)$_2$As$_2$ (Co-Ba122)\cite{Meingast2012} or BaFe$_2$(As,P)$_2$(P-Ba122)\cite{Boehmer2012}. The very small anomaly in the volume average of the thermal expansion indicates that $T_s$ does not couple strongly to hydrostatic pressure. No distinct second, potentially magnetic, phase transition is observed below $T_s$. At $T_c=7.75$ K, the discontinuity in the in-plane thermal-expansion coefficients $\Delta\alpha_i$ (kink in $\Delta L_i(T)$) clearly confirms a sharp, bulk superconducting transition. $\Delta\alpha_i$ is related to the uniaxial pressure derivative of $T_c$ via the Ehrenfest relationship 
\begin{equation}
\frac{dT_c}{dp_i}=\frac{V_m\Delta\alpha_i}{\Delta C_p/T_c}.\label{eq:Ehrenfest}
\end{equation}
$V_m=23.34$ cm$^3$/mol is the molar volume and $\Delta C_p/T_c=9.45(30)$ mJ mol$^{-1}$K$^{-2}$ is the specific heat jump, which we take from Reference \onlinecite{Lin2011}. We thus obtain $\frac{dT_c}{dp_{twin}}=2.6(3)$K/GPa for the in-plane average, $\frac{dT_c}{dp_a}=2.2(5)$ K/GPa, $\frac{dT_c}{dp_b}=3.1(1.1)$ K/GPa and $\frac{dT_c}{dp_c}=0(0.5)$K/GPa. The hydrostatic pressure derivative of $T_c$ is simply given by the sum of the uniaxial components  $\frac{dT_c}{dp_{v}}=\frac{dT_c}{dp_{a}}+\frac{dT_c}{dp_{b}}+\frac{dT_c}{dp_c}=5.3(1.2)$ K GPa$^{-1}$ and is in good agreement with the initial slope of direct measurements, which yield $\frac{dT_c}{dp_{v}}=6-7$ K GPa$^{-1}$\cite{Masaki2009,Bendele2012}. It is clear from our results that the comparatively large $\frac{dT_c}{dp_{v}}$ in FeSe arises not from particularly large uniaxial components but from lack of their cancellation. The in-plane derivatives are comparable in size to slightly overdoped Co-Ba122\cite{Budko2009,Hardy2009}. However, in Co-Ba122, in-plane and $c$-axis pressure derivatives have opposite signs and largely cancel in the hydrostatic average (with the negative $p_c$ derivative slightly prevailing)\cite{Budko2009,Meingast2012}, while the $p_c$ derivative is approximately zero in FeSe. The structural tuning parameter of the 122-systems is the $c/a$ ratio\cite{Hardy2009,Meingast2012,Boehmer2012}. In FeSe instead, the in-plane distance alone appears to be the tuning parameter, which couples strongly to hydrostatic pressure. 

Basically the same picture emerges when considering the uniaxial strain derivatives of $T_c$, $\frac{dT_c}{d\varepsilon_j}=\sum_{i} c_{ij} \frac{dT_c}{dp_i}$, which can be calculated if the set of elastic constants $c_{ij}$ $(i,j=1-3)$ is known. For tetragonal FeSe, these constants have been calculated using DFT\cite{Chandra2010} to $c_{11}=95.2$ GPa, $c_{12}=48.8$ GPa,  $c_{13}=13.9$ GPa  and $c_{33}=39.5$ GPa\cite{noteelconstantsFeSe}.
Using  the above $c_{ij}$ values (allowing for an error of 10\%) we find $\frac{dT_c}{d\varepsilon_a}=-399(177)$ K, $\frac{dT_c}{d\varepsilon_b}=-358(142)$ K and $\frac{dT_c}{d\varepsilon_c}=-73(49)$ K. It is evident that $T_c$ depends sensitively on the in-plane lengths. The corresponding uniaxial Grüneisen parameter is $\frac{d\ln T_c}{d\varepsilon_{ab}}=-49(12)$, \textit{i. e.} shrinking $(a+b)/2$ by 1\% increases $T_c$ by $\sim50\%$. Note that $a$ and $b$ decrease by $\sim1.5$\% between ambient pressure and 7 GPa, where the highest $T_c$ is reached\cite{Margadonna2009}. $T_c$ is five times less sensitive to changes of the $c$-axis length ($\frac{d\ln T_c}{d\varepsilon_c}=-9(6)$). 
It is striking that the $c$-axis length has such a small effect on $T_c$, especially since the Se height (i.e. the $c$-axis length times the internal  $z$-parameter) was found to correlate closely with $T_c$ \cite{Mizuguchi2010}. The $z$-parameter however, may depend in a complicated manner both on in-plane and $c$-axis lengths. For a detailed investigation of its relation to uniaxial pressure effects, this dependence would have to be established.

\begin{figure}[tb]
\includegraphics[width=8cm]{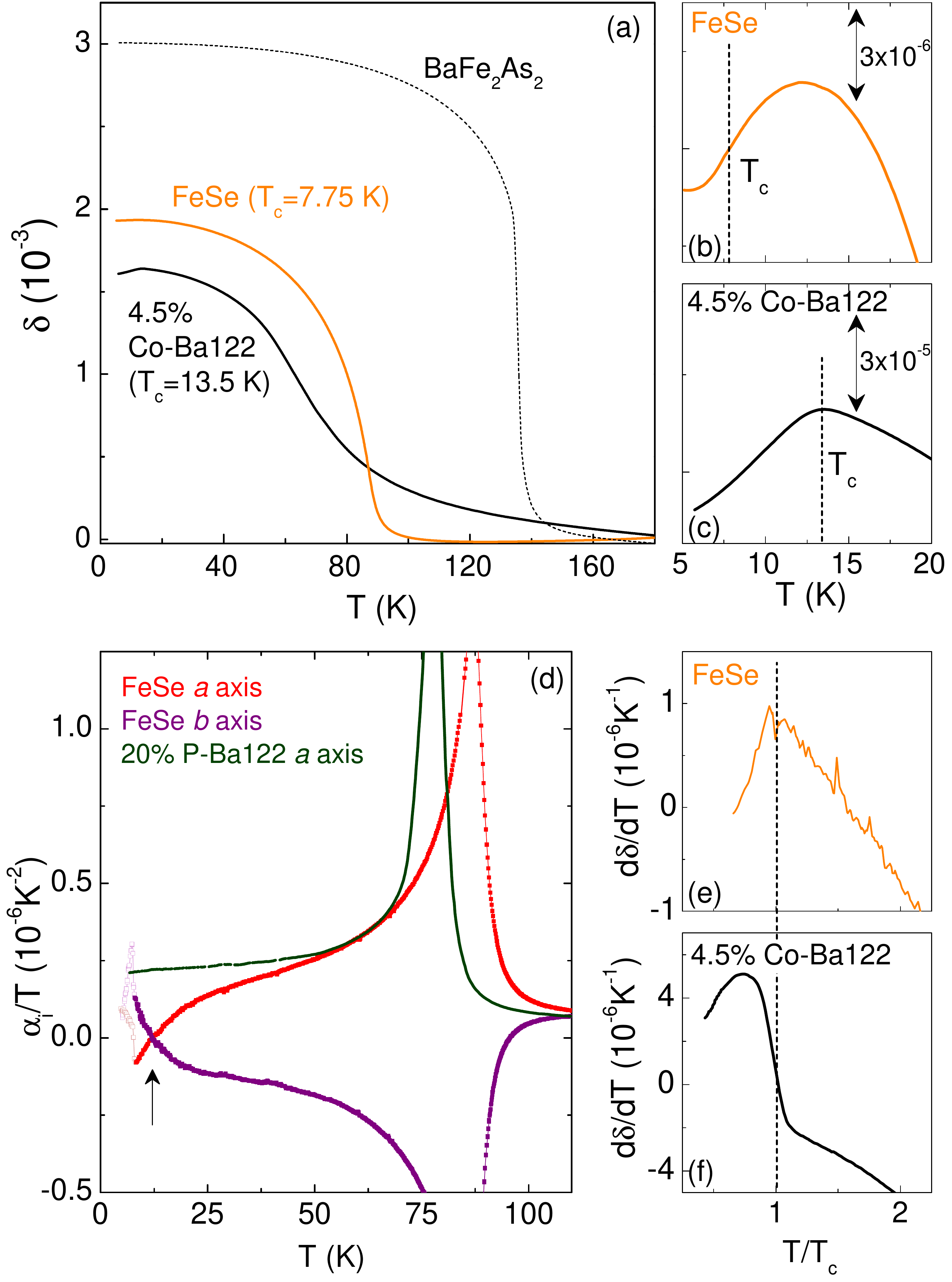}
\caption{(color online) (a) Orthorhombic order parameter $\delta=|b-a|/2a_0$ of FeSe (orange line), undoped Ba122 (dashed black line) and underdoped Co-Ba122 (black line) computed from thermal-expansion measurements. (b) and (c) show $\delta(T)$ of FeSe and Co-Ba122, respectively, below 20 K on a magnified scale. Note that the scale in (b) is ten times smaller than in (c). (d) $\alpha_{a,b}/T$ of FeSe (red and purple lines) and $\alpha_{a}/T$ of underdoped P-Ba122 (dark green line). For clarity, data below $T_c$ are shown in a lighter color. The arrow points at the crossing of the normal-state in-plane thermal-expansion coefficients of FeSe. (e) and (f) show $\frac{d\delta}{dT}$ of FeSe and Co-Ba122, respectively, close to $T_c$. }
\label{fig:3}
\end{figure}

Fig. \ref{fig:3} presents our results concerning the interplay between orthorhombicity, magnetic fluctuations and superconductivity in FeSe. Fig. \ref{fig:3} (a) shows the orthorhombic order parameter $\delta=|a-b|/(a+b)\approx|a-b|/2a_0$ of FeSe, undoped Ba122 and underdoped Co-Ba122 (4.5\% Co-content), all computed from our thermal-expansion measurements. Although no magnetic transition is found, $\delta(T=0)$ of FeSe is of similar magnitude as in the 122-systems. While $\delta(T)$ is reduced below $T_c$ in the 4.5\%Co-Ba122, there is no discernable feature in $\delta(T)$ of FeSe at $T_c$ (Fig. \ref{fig:3} (b,c)). Curiously, however, $\delta(T)$ has a weak maximum at $\sim 12$ K, which we will discuss before we address the response of $\delta(T)$ at $T_c$ in more detail. 

The maximum of $\delta(T)$ is caused by the sign change of both $\alpha_a(T)$ and $\alpha_b(T)$ at $\sim12$ K (see inset Fig. \ref{fig:2}(e)), which points to an additional low-temperature contribution to the thermal expansion. 
This contribution, which appears to be diverging down to $T_c$, becomes evident in a plot of $\frac{\alpha_{a,b}}{T}$ (Fig. \ref{fig:3} (d)). For a Fermi-liquid, one expects a constant $\frac{\alpha}{T}$-term at low temperatures, which is directly related to the uniaxial pressure derivative of the Sommerfeld coefficient, as seen for P-Ba122 with 20\% P-content\cite{Boehmer2012} (Fig. \ref{fig:3}(d)). 
%We therefore consider in Fig. \ref{fig:3} (d) $\frac{\alpha_{a,b}}{T}$, which for a Fermi-liquid is temperature independent and equals the negative pressure derivative of the Sommerfeld coefficient $\gamma$ (excepting a lattice contribution). Fig. \ref{fig:3} (d) also shows $\frac{\alpha_a}{T}$ of underdoped, non-superconducting, P-Ba122 (20\% P-content) which clearly exhibits Fermi-liquid behavior in the orthorhombic and magnetically-ordered state\cite{Boehmer2012}. 
The non-Fermi-liquid character of the thermal-expansion of FeSe becomes apparent in this comparison. A new energy scale with a negative (positive) contribution to $\frac{\alpha_a}{T}$ ($\frac{\alpha_b}{T}$) emerges below $\sim30$ K and causes these coefficients to diverge. Note that neither the Knight-shift, nor specific heat have previously shown indications of non-Fermi-liquid behavior \cite{Lin2011,Imai2009}. However, thermal-expansion is expected to be an especially sensitive probe for locating such non-Fermi-liquid behavior\cite{Garst2005} and enhanced low-temperature spin fluctuations have reavealed that FeSe is on the brink of a magnetic phase transition near $T=0$ \cite{Imai2009}. Our observation of non-Fermi-liquid behavior is probably related to this quantum critical point and the new energy scale may be linked to the low-temperature spin fluctuations. It is then, however, curious that these fluctuations cause a reduction of the orthorhombic order parameter of FeSe. In the 122-pnictides, in contrast, the onset of magnetism enhances $\delta(T)$\cite{Kim2011}, suggesting that the relation between magnetism and orthorhombicity may be different in FeSe. 

The very small $ab$-plane anisotropy of $\Delta\alpha_i$ of FeSe at $T_c$ implies that $\delta(T)$ has no more than a tiny anomaly at $T_c$, as can indeed be seen in Fig. \ref{fig:3} (b). Only a small kink in the temperature derivative $\frac{d\delta}{dT}$ (Fig. \ref{fig:3}(d)) is observed, which is such that $\delta$ will have a tendency to be somewhat larger in the superconducting state than in the normal state. In particular, there is no indication of competition between orthorhombicity and superconductivity in FeSe. Such a competitive coupling was clearly observed in underdoped Co-Ba122 (see Fig. \ref{fig:3}(c) and Ref. \onlinecite{Nandi2010}) and other doped Ba122-compounds\cite{Avci2011} and results in a reduction of $\delta$ below $T_c$, or equivalently in a positive anomaly in $\frac{d\delta}{dT}$ (see Fig. \ref{fig:3}(f)). This coupling was proposed to arise, ultimately, from the competition between magnetism and superconductivity\cite{Nandi2010}. In this scenario, magnetic spin fluctuations, which give rise to ``nematic'' order that, in turn, induces an orthorhombic distortion via magneto-elastic coupling, are weakened by the onset of superconductivity. It is an important question which part of this scenario is not valid for FeSe. Our results suggest that either superconductivity does not interact strongly with spin fluctuations or that the structural transition has a non-magnetic origin. 

In conclusion, high-resolution thermal-expansion measurements of vapor-grown $\beta$-FeSe have revealed a number of unusual properties, which suggest that FeSe is not a typical iron-based superconductor. The structural tuning parameter of FeSe is the in-plane area, compared to the $c/a$ ratio in 122-systems, which explains why $T_c$ couples so strongly to hydrostatic pressure. In underdoped 122 pnictides, magnetism and orthorhombicity are cooperative and compete with superconductivity. In FeSe, on the other hand, $T_c$ and magnetism both increase under hydrostatic pressure, suggesting that they act cooperatively. Further, superconductivity does not compete with orthorhombicity, which raises the question of the origin of the orthorhombic phase transition. Also, we show that a new energy scale, presumably associated with spin fluctuations, emerges at low temperature and slightly reduces the orthorhombic distortion below 12 K. Further studies on the nature of these different phases, which exhibit such an unusual interplay, will be of great interest.  

We wish to thank J. Schmalian and R. M. Fernandes for discussions. This work was supported by the DFG under the priority program SPP1458.

%\bibliography{D:/FeSe/referencespnictidesall}

%\bibliography{C:/Users/boehmer/Documents/reports/referencespnictidesall}

\begin{thebibliography}{29}
\bibitem [{not({\natexlab{a}})}]{notenamingab}%
  \BibitemOpen
  \href@noop {} {}\ \bibinfo {note} {{Following} conventions,
  we call the shorter orthorhombic in-plane axis $a$, opposed to the prevailing
  custom in 122-pnictides.}\BibitemShut {Stop}%
\end{thebibliography}
%\bibliographystyle{apsrev4-1} 

\end{document}